\documentclass[10pt,twoside,twocolumn,fleqn]{article}

\usepackage[
  paperwidth=595.276bp,
  paperheight=790.866bp,
  left=51.0bp,
  right=48.0bp,
  top=52.0bp,
  bottom=47.0bp,
  headheight=14bp,
  headsep=8bp,
  footskip=23bp
]{geometry}
\usepackage[T1]{fontenc}
\usepackage[utf8]{inputenc}
\usepackage{textcomp}
\IfFileExists{newtxtext.sty}{%
  \usepackage{newtxtext,newtxmath}%
}{%
  \usepackage{mathptmx}%
}
\usepackage[scaled=0.94]{helvet}
\usepackage{microtype}

\usepackage{amsmath,amsfonts}

\usepackage{graphicx}
\usepackage[table]{xcolor}
\usepackage{booktabs}
\usepackage{array}
\usepackage{tabularx}
\usepackage{enumitem}
\usepackage{caption}
\usepackage{fancyhdr}
\usepackage{fontawesome5}
\usepackage{titlesec}
\usepackage[round,authoryear]{natbib}
\usepackage{xurl}
\usepackage{bm}
\usepackage{siunitx}
\usepackage{hyperref}
\IfFileExists{orcidlink.sty}{\usepackage{orcidlink}}{\newcommand{\orcidlink}[1]{}}

\definecolor{RAgray}{gray}{0.82}
\definecolor{RAlightgray}{gray}{0.92}
\definecolor{RAblue}{RGB}{0,0,180}
\hypersetup{
  colorlinks=true,
  linkcolor=RAblue,
  citecolor=RAblue,
  urlcolor=RAblue,
  pdfauthor={First A. Author},
  pdftitle={Final-format template}
}
\graphicspath{{figures/}}

\newcommand{\RAjournal}{ }
\newcommand{\RAyear}{ }
\newcommand{\RAvolume}{ }
\newcommand{\RApages}{ }
\newcommand{\RAdoi}{XX.XXXX/sXXXXX-XXX-XXXXX-X}
\newcommand{\RAtype}{TECHNICAL NOTE}
\newcommand{\RAtitle}{PocketCaBER and PocketDoS: Low-cost open-source tools for teaching and learning advanced topics in fluid mechanics}
\newcommand{\RAauthors}{Zhaofeng Peng$^\dagger$\textsuperscript{1}\,\orcidlink{0009-0009-9134-3001} \;\textperiodcentered\; Lucas Warwaruk$^\dagger$\textsuperscript{2}\,\orcidlink{0000-0003-0676-5971} \;\textperiodcentered\; Thomas Livesay\textsuperscript{3} \;\textperiodcentered\; Benjamin M. Yavitt\textsuperscript{4} \;\textperiodcentered\; Randy H. Ewoldt\textsuperscript{3} \;\textperiodcentered\; Gareth H. McKinley\textsuperscript{2}\,\orcidlink{0000-0001-8323-2779} \;\textperiodcentered\; Laurel Kroo\textsuperscript{1}\,\orcidlink{0000-0001-7860-2632}}
\newcommand{\RAdates}{Dated: \today}
\newcommand{\RAcopyright}{\copyright\ The Author(s) 2026}
\newcommand{\RAkeywords}{CaBER \;\textperiodcentered\; DoS \;\textperiodcentered\; Frugal science \;\textperiodcentered\; Rheometry}
\newcommand{\RAabstract}{We describe two open-source, 3D-printable, flexure-based tools for the quantitative measurement of extensional properties of viscoelastic fluids. These low-cost, portable, and scalable devices (which we have termed ``PocketCaBER'' and ``PocketDoS'') are particularly applicable for use in the field and in graduate-level teaching environments due to their low cost, printability on hobby 3D printers, compatibility with cell phone cameras, portability and user-friendly operation. We characterize and benchmark each device's performance against its lab-equivalent counterpart and provide downloadable STL files for rapid fabrication. We discuss experimental limitations of these devices compared with their bench-top counterparts. Finally, we illustrate the use of such tools in facilitating student engagement in polymer science and complex fluids classes---specifically, how progress in learning goals can be uniquely and effectively accelerated by providing the necessary rheological instruments directly to each individual student (especially for advanced modules such as nonlinear extensional rheology). By giving students personal, indefinite access to laboratory-level instrumentation through these open-source frugal science tools, we discuss our efforts to expand participation and engagement within the field of nonlinear rheology.}
\newcommand{\RAshorthead}{\RAjournal\ \RAyear \RAvolume\RApages}

\newcommand{\RAauthorblock}{%
\faIcon[regular]{envelope}\ Laurel Kroo\\
\href{mailto:lkroo@umass.edu}{lkroo@umass.edu}\par\smallskip
\textsuperscript{1}\ Department of Polymer Science and Engineering, University of Massachusetts Amherst, Amherst, MA 01003, USA\par
\textsuperscript{2}\ Department of Mechanical Engineering, Massachusetts Institute of Technology, Cambridge, MA 02139, USA\par
\textsuperscript{3}\ Department of Mechanical Science and Engineering, University of Illinois Urbana-Champaign, Urbana, IL 61801, USA\par
\textsuperscript{4}\ Department of Chemical Engineering, University of Cincinnati, Cincinnati, OH 45221, USA\par
\textsuperscript{$\dagger$}\ These authors contributed equally to this work}%

\fancypagestyle{RAstyle}{%
  \fancyhf{}%
  \fancyhead[LE]{\sffamily\footnotesize\thepage}%
  \fancyhead[RO]{\sffamily\footnotesize\thepage}%
  \fancyhead[LO]{\sffamily\footnotesize\RAshorthead}%
  \fancyhead[RE]{\sffamily\footnotesize\RAshorthead}%
  \fancyfoot[LE,RO]{\sffamily\footnotesize }%
}
\fancypagestyle{RAfirst}{%
  \fancyhf{}%
  \fancyfoot[R]{\sffamily\footnotesize }%
}
\titleformat{\section}
  {\sffamily\bfseries\fontsize{13.0}{15.5}\selectfont}
  {}{0pt}{}
\titlespacing*{\section}{0pt}{16pt plus 3pt minus 2pt}{7pt}
\titleformat{\subsection}
  {\sffamily\bfseries\fontsize{10.8}{12.8}\selectfont}
  {}{0pt}{}
\titlespacing*{\subsection}{0pt}{12pt plus 2pt minus 1pt}{5pt}
\titleformat{\subsubsection}
  {\sffamily\bfseries\fontsize{10.0}{12.0}\selectfont}
  {}{0pt}{}
\titlespacing*{\subsubsection}{0pt}{10pt plus 2pt minus 1pt}{4pt}

\setlist{nosep,leftmargin=*}

\newcommand{\makeRAfrontmatter}{%
  \thispagestyle{RAfirst}%
  \begingroup
  \vspace*{-18pt}%
  \noindent{
  }\par
  \vspace{7pt}
  \noindent\rule{\textwidth}{0.65pt}\par\vspace{0pt}
  \noindent\colorbox{RAgray}{\makebox[0.48\textwidth][l]{\hspace{6pt}\sffamily\bfseries\fontsize{8.6}{10.2}\selectfont\RAtype}}\par
  \vspace{26pt}
  \noindent{\sffamily\bfseries\fontsize{17.2}{20.0}\selectfont \RAtitle\par}
  \vspace{17pt}
  \noindent{\sffamily\bfseries\fontsize{9.4}{11.2}\selectfont \RAauthors\par}
  \vspace{19pt}
  \noindent{\sffamily\fontsize{8.2}{9.9}\selectfont \RAdates\\[-1pt]\RAcopyright\par}
  \vspace{20pt}
  \noindent{\sffamily\bfseries\fontsize{9.8}{11.0}\selectfont Abstract}\par
  \vspace{2pt}
  \noindent{\fontsize{9.2}{11.0}\selectfont \RAabstract\par}
  \vspace{11pt}
  \noindent{\fontsize{9.2}{11.0}\selectfont {\sffamily\bfseries Keywords}\quad \RAkeywords\par}
  \vspace{25pt}
  \endgroup
}

\makeatletter
\newcommand{\makeRAauthorfootnote}{%
  \begingroup
  \renewcommand{\thefootnote}{}%
  \renewcommand{\@makefntext}[1]{%
    \parindent=0pt
    \noindent
    ##1%
  }%
  \footnotetext[0]{%
    \fontsize{8.0}{9.2}\selectfont
    \RAauthorblock
  }%
  \endgroup
}
\makeatother

\begin{document}

\twocolumn[\makeRAfrontmatter]
\thispagestyle{RAfirst}
\makeRAauthorfootnote

\begin{figure*}[ht!]
    \centering
    \includegraphics[width=\textwidth]{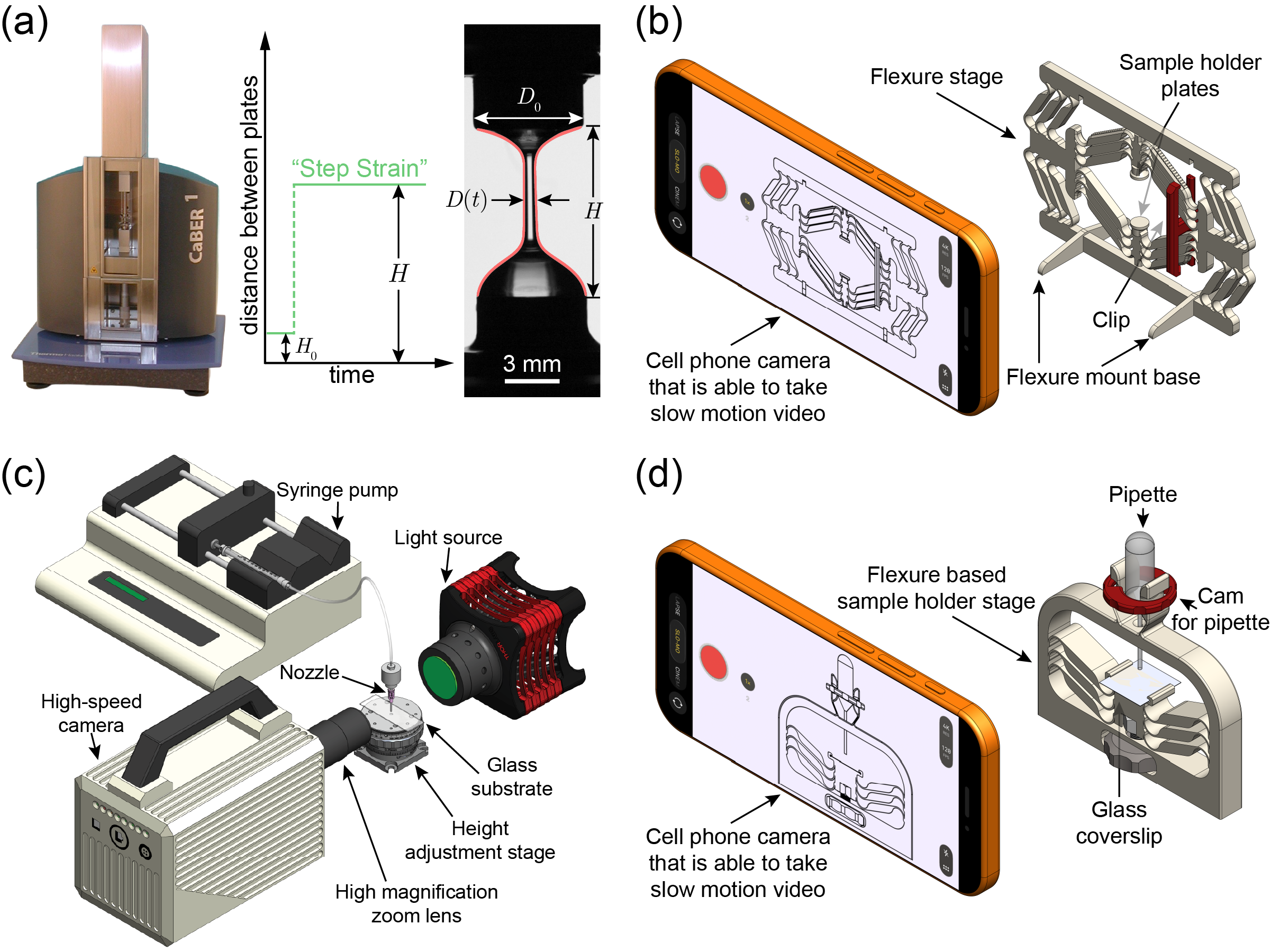}
    \caption{\textbf{Comparison of research-grade “gold standard” equipment ((a) and (c)) and pocket devices ((b) and (d))} (a) A traditional capillary breakup extensional rheometer (CaBER) is shown (designed by Cambridge Polymer Group and sold by Thermo Haake). It employs a step strain mechanism to probe the extensional rheological properties of the sample, which can be inferred from the diameter evolution $D(t)$ of the formed liquid filament. The rightmost panel is a sample image taken from the high-speed camera for a PIB solution, in which the red contour line represents the air-liquid interface of the liquid sample, with $D_0$ as the sample plate diameter, and $H$ the final separation distance between two sample plates. (b) Shown here is the PocketCaBER design, which leverages a flexure mechanism to achieve the step strain. A cell phone is used to replace the high-speed camera. A macro lens is attached to the cell phone camera when high magnification is needed. A light source can be added to the back of the device for a better contrast in detecting the edge of the liquid sample. (c) The dripping-onto-substrate (DoS) technique is shown here (included with permission from \citet{warwaruk2026}, licensed under a Creative Commons Attribution (CC BY) license---https://creativecommons.org/licenses/by/4.0/). (d) The PocketDoS design, similar to PocketCaBER, is shown here. It requires a cell phone camera with slow-motion mode.}
    \label{Fig:0}
\end{figure*}

\noindent Low-cost, accessible scientific tool development is a crucial area of engineering education and enables broad participation in global scientific discovery. For example, following the invention of Foldscope \citep{Cybulski2014}, a \$1.00 USD flexure-based paper microscope, the global amateur microscopy community expanded significantly---forming the largest-to-date active central online microscopy platform \citep{microCosmos} where amateur scientists share discoveries of real-world materials and biological systems. This pattern is common for scientific instrumentation: when an advanced tool rapidly becomes orders of magnitude more accessible, scientific progress follows within that discipline as the size and diversity of perspective of the engaged community expands. Even beyond  curiosity-driven science, these tools can also rapidly enable critical emerging technologies in fields such as global healthcare diagnostics \citep{korir2015punch,Bhamla2017} or environmental surveillance \citep{mukundarajan2017using,kumar2021microfluidic}, expanding broader impacts and bridging the conventional gap between teaching and research. These examples elucidate the effect of expanding scientific access beyond the traditional laboratory: direct technical and social impact follows when experiments are available to everyone with curiosity and initiative.

Recurring core design principles are notable in the field of frugal science which dramatically improve accessibility of scientific instrumentation. For example, replacing precision stages with flexures (as we do in this present work) was also the key design innovation that Foldscope employed \citep{Cybulski2014} to minimize the cost associated with integrating micromechanical optical stages. Other common design methods include replacing scientific cameras with modern cell phone technology \citep{breslauer2009mobile}, reducing aberrations in optical paths by minimizing focal distances and/or using inverted cell phone camera lens modules \citep{switz2014low}, among many other common design approaches. This work leverages several of these recurring principles.   

In the field of rheology, many efforts have been made consistent with these frugal science approaches---to advance rheological techniques in the field, in the classroom and across industries. State-of-the-art shear rheometers that measure the properties of complex fluids (viscosity, first normal stress difference, \textit{etc.}) are at the time of this publication approximately in the range of \$100,000--\$300,000 USD and are capable of precision measurements (for example, the TA Instruments ARES-G3 has a stated torque resolution of \qty{1}{nN\cdot m}). 
Significant efforts \citep{antignard2021portable,erni2024low} have been made to create DIY devices from off-the-shelf electro-mechanical components that are lower cost, by sacrificing some accuracy and precision.

The field of extensional rheometry---which is central to polymer science, food science and industrial processing---has also seen numerous developments in frugal instrumentation. Notably, the dripping-onto-substrate (DoS) method \citep{Dinic2017} was developed to study extensional properties of fluids with lower values of the dynamic viscosity than could be considered by previous extensional methods, such as capillary breakup extensional rheometry (CaBER) \citep{warwaruk2026}. But DoS was also inspired by the relative simplicity of the technique compared with other methods (\textit{i.e.}, there is no need for fast-moving actuators or precision positioning stages), which reduced the cost of instrumentation and expanded access. 

In this same spirit, \citet{Marshall2017} proposed a clothespin-based handheld device as a method to mimic CaBER, at exceptionally low cost---departing from precision electronics and leveraging approximate observations without the strict need for a controlled laboratory environment. This device, however, had some limitations in that it imparts some rotational kinematics (\textit{i.e.}, angular displacement) to the sample and user-defined variable rate of imposed step-strain. Recent work has also brought to light sample size and gravity sensitivity of some of these extensional techniques \citep{Gaillard2024,hu2025revealing} leading to recent development leveraging liquid-in-liquid electro-mechanical setups to mitigate gravitational effects \citep{te2026portable}. Other recent open-source efforts by \citet{Calabrese2026Stringimeter} also using electro-mechanical components have shown excellent instrument performance compared with the state-of-the-art. However, frugal science teaching tools in the space of extensional rheometry are still lacking---especially instrumentation designs that can be fabricated in bulk for large classrooms. Moreover, it is necessary for these instruments to be sufficiently inexpensive (\textit{e.g.}, $\ll 1$ USD) such that each student/user can independently ask scientific questions and keep the instrument for use outside of the traditional classroom or teaching environment. In this study, we approach this challenge via the design of a flexure-based, solely mechanical linear mechanism to impart the classical kinematics of capillary breakup rheometry and dripping-onto-substrate measurement methods on fluid samples. The step-strain rate of the stage is significantly faster than recently developed electro-mechanical methods, and in this work we show that it is  similarly effective at repeatable measurement capabilities (but critically, at 1--2 orders of magnitude lower in cost).

Bridging research and teaching in this way also requires designs that are operationally simple enough to focus on the science rather than the construction and tuning of the measurement instrument. Indeed, the accessibility of these teaching tools also goes beyond cost-minimization of instrumentation; \citet{Hossain2024} developed the idea of ``protorheology'', in which simple rheological observations from toy-experiments can inform meaningful (and surprisingly accurate) conceptual characterization of the dynamic properties of complex fluids. This approach has proven to be particularly powerful in teaching contexts, where exact measurements are less important than ease-of-use or rapid observation of trends. In classroom environments, the demonstration of a surprising phenomena is oftentimes more important than its precise quantification. 

In this work, we have tried to bridge these two domains of frugal science and protorheology by designing and extensively testing two new instruments for studying the extensional properties of complex fluids, which leverage recent advances in 3D printing and flexure design to replace precision/high-speed stages. These instruments cost less to fabricate than 50 cents each, require no electronic components besides a cell phone camera, and have been tested extensively at scale to assess pedagogical impact.

\section*{Brief Overview: Accessible Extensional Rheometry}

Extensional rheometry is a technique for probing the mechanical properties of a sample, such as the transient extensional viscosity and the relaxation time. It involves the formation of a sample liquid filament, which thins gradually until it breaks. The time evolution of the rapidly-thinning filament diameter encodes the sample properties; thus, a high-speed camera is often needed to film the (often millisecond-scale) process, adding to the overall cost of the equipment. Based on how the filament is formed, extensional rheometry can be further divided into multiple different methods, including capillary breakup extensional rheometry (CaBER) and dripping-onto-substrate (DoS) method, \textit{etc.} For CaBER, the liquid filament is formed by rapidly separating two sample plates, between which the sample is constrained. This technique often involves a precision linear stage, leading to expensive and delicate laboratory equipment (see the leftmost figure in Fig. \ref{Fig:0}(a)). For DoS, the liquid filament is formed by dripping a sample droplet onto a substrate. This method uses a syringe pump for precision dispensing the droplet, resulting in a similarly delicate and complex instrument design (see Fig. \ref{Fig:0}(c), \citet{warwaruk2026}).

 CaBER and DoS methods probe sample properties associated with shear-free uniaxial extensional flows: the flow type within an idealized thinning liquid filament. In such flows, the deformation (of a fluid element) is often described by the elongation ratio $\lambda$, which is the current length of a material element $L$ divided by its original length $L_0$, \textit{i.e.}, $\lambda = L/L_0$. This sets the Hencky strain, ${\varepsilon}$, which is the natural logarithm of the elongation ratio of a material element: $\varepsilon = \ln{\lambda}$. The Hencky strain rate, $\dot{\varepsilon}$, is the time derivative of this Hencky strain. This extensional flow leads to a time-dependent filament radius. For a viscoelastic liquid sample, such as a dilute polymer solution, the time evolution of the formed filament can be divided into three distinct regimes, namely inertio-capillary (IC) regime/visco-capillary (VC) regime (delineation is discussed elsewhere but is determined by the ratio between viscous and inertial forces, \textit{i.e.}, Ohnesorge number, see: \citet{mckinley2011wolfgang} and \citet{warwaruk2026}), elasto-capillary (EC) regime, and finite extensibility/breakup regime. The EC regime results from the balance between elastic and capillary stresses in the liquid filament, which encompasses rich information about the viscoelastic properties of the sample. Among these properties, the relaxation time, $\tau$, describes how rapidly elastic stresses dissipate in a stress-relaxation experiment. 


\citet{entov1997effect} showed that the product of the Hencky strain rate and the relaxation time, also known as the Weissenberg number (Wi), is pinned at $\mathrm{Wi}=\dot{\varepsilon}\tau=2/3$ in the EC regime (using the assumption of an Oldroyd-B fluid). This property of the EC regime arises from a force balance between surface tension and elasticity, constraining the Hencky strain rate. They used thinning dynamics of the filament to infer relaxation time from the EC regime, as is now a common approach to the field of extensional rheometry. Here we show a simplified version of this derivation using the upper-convected Maxwell (UCM) model: a common constitutive model used to describe the behavior of polymer melts in extensional and nonlinear flows. 

The UCM model gives us a relationship between the stress tensor and the rate-of-strain tensor for a prototypical viscoelastic fluid: 
\begin{equation}
    \bm{\sigma}+\tau\overset{\triangledown}{\bm{\sigma}}=\eta_p\dot{\bm{\gamma}},
    \label{UCM}
\end{equation}
where $\bm{\sigma}$ is the polymeric stress tensor, $\eta_p$ is the polymeric contribution to the viscosity, and $\dot{\bm{\gamma}}=\nabla\bm{v}+(\nabla\bm{v})^\mathrm{T}$ is the rate-of-strain tensor, with $\nabla\bm{v}$ as the velocity gradient of the flow field. Here, $\overset{\triangledown}{\bm{\sigma}}=\frac{D\bm{\sigma}}{Dt}-\left[(\nabla\bm{v})^\mathrm{T}\cdot\bm{\sigma}+\bm{\sigma}\cdot(\nabla\bm{v})\right]$ denotes the upper-convected time derivative of the stress tensor. In the EC regime, using a cylindrical coordinate system, a uniaxial stretching flow on a uniform cylindrical filament  with a velocity field $v_r=-\frac{\dot{\varepsilon}}{2}r$ and $v_z=\dot{\varepsilon}z$ will result in an axial stress evolution from Eq. \eqref{UCM} given by:
\begin{equation}
    \frac{\mathrm{d}}{\mathrm{d}t}\sigma_{zz}=\left(2\dot{\varepsilon}-\frac{1}{\tau}\right)\sigma_{zz}+\dfrac{2\eta_p}{\tau}\dot{\varepsilon}.
    \label{beforeSimplification_z_component}
\end{equation}
We observe that the terms multiplied by $\sigma_{zz}$ will rapidly exceed $\eta_p/\tau$ at finite time because the polymeric axial stress will grow unbounded, whereas the last term is effectively a steady-state offset. If we neglect this offset term, our constitutive equation in the axial direction becomes simply: 
\begin{equation}
    \frac{\mathrm{d}}{\mathrm{d}t}\sigma_{zz}=\left(2\dot{\varepsilon}-\frac{1}{\tau}\right)\sigma_{zz}.
    \label{z_component}
\end{equation}


A separate kinematic constraint on the system arises from the assumption that the extra stresses have the relationship $|\sigma_{rr}| \ll |\sigma_{zz}|$ because the polymers are oriented in the axial direction. In this simplified model, we neglect the solvent viscosity. When we consider the \textit{total} stress (\textit{i.e.}, $\bm{T} = -p\bm{I}+\bm{\sigma}$), it is evident that the radial component ($T_{rr}$) must balance the capillary stress across the boundary, while the axial component ($T_{zz}$) approaches zero because the top and bottom of the filament are attached to a solid boundary (or as described originally by \citet{entov1997effect}, attached to ``large stagnant drops on stationary end plates'').
These nuanced set of assumptions set up a constraint between the axial polymeric stress in the sample and surface tension, given by: 
\begin{equation}
    \sigma_{zz}=\frac{\Gamma}{R},
    \label{B.C.}
\end{equation}
where $\Gamma$ is the surface tension and $R$ is the radius of the liquid filament. 

Another kinematic boundary condition arises from the time derivative of the filament radius equaling the radial velocity at the free surface:  
\begin{equation}
    \frac{\mathrm{d}R}{\mathrm{d}t}=-\frac{\dot{\varepsilon}}{2}R. 
    \label{incompressibility}
\end{equation}

With these three independent descriptors of the system response (Eqs. \eqref{z_component}, \eqref{B.C.} and \eqref{incompressibility}), we can describe the relationship between the material properties (\textit{i.e.}, $\tau$) and the system-determined rate of deformation ($\dot{\varepsilon}$). Differentiating both sides of Eq. \eqref{B.C.} with respect to time and subsequently equating the right-hand side with Eq. \eqref{incompressibility} yields $\dot{\sigma}_{zz}=\frac{\dot{\varepsilon}}{2}\sigma_{zz}$, which, when combined with Eq. \eqref{z_component}, leads to:
\begin{equation}
    \mathrm{Wi}=\dot{\varepsilon}\tau=\frac{2}{3}.
    \label{Wi}
\end{equation}
The implication of Eq. \eqref{Wi} is consequential: capillarity selects a particular stretching rate in thinning filaments, and the  rate is determined by the viscoelasticity of the sample (\textit{e.g.}, the fluid's relaxation time).   

By combining Eqs. \eqref{incompressibility} and \eqref{Wi} and then integrating with respect to time leads to
\begin{equation}
    R\propto\exp\left(-\frac{t}{3\tau}\right),
    \label{radius}
\end{equation}
demonstrating that the expected filament radius will decay in an exponential manner in the elasto-capillary regime for a prototypical viscoelastic fluid. A leading proportionality constant which is dependent on the surface tension of the fluid is omitted here (but discussed further in Table \ref{tab:discussion}). 

This derivation informs two key theory principles. First, we have shown (Eq. \eqref{Wi}) that the rate of stretching is not a controlled parameter in capillary-driven filament thinning rheometry; it is \textit{selected} by the material itself: viscoelastic fluids with small but non-zero relaxation times will generate large Hencky strain rates in the EC regime; viscoelastic fluids with larger relaxation times (more elastic/solid-like) will generate smaller Hencky strain rates in capillary-driven thinning. Second, we expect the radius of a thinning viscoelastic filament to scale as a decreasing exponential function (Eq. \eqref{radius}), where the slope of the radial evolution (on a semi-logarithmic axis) reflects the relaxation time of the material. We qualify these two principles with the statement that in this section (and also in \citet{entov1997effect}), a specific relationship was assumed between stress and rate-of-strain (UCM model) that does not necessarily apply to all complex fluids. It is, however, representative of the key dynamics for many classes of polymer melts and polymeric solutions. These findings are also only relevant to the EC regime.  

\begin{figure*}[ht!]
    \centering
    \includegraphics[width=\textwidth]{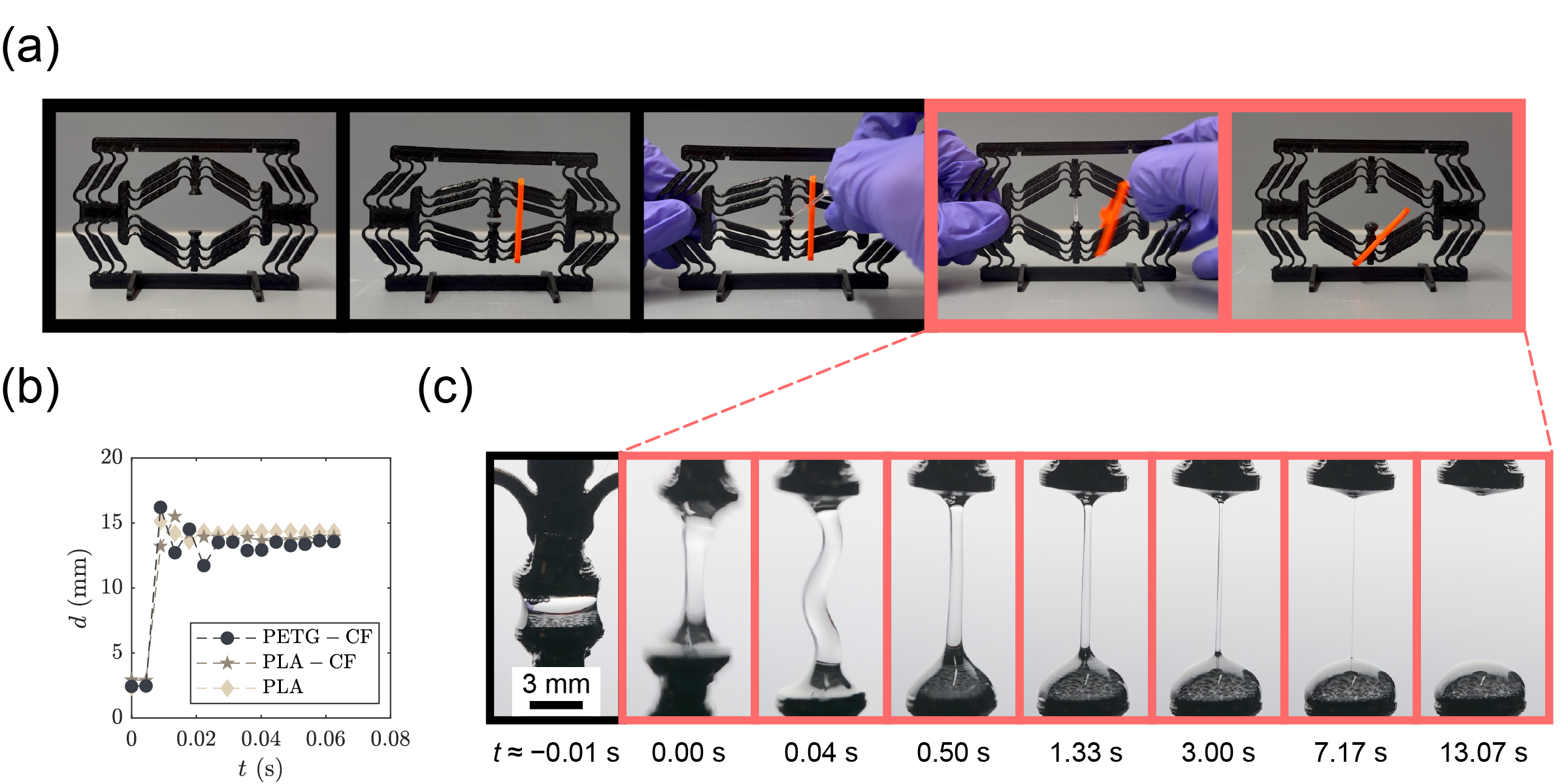}
    \caption{\textbf{Design of PocketCaBER} (a) The experimental protocol of PocketCaBER is shown. Before an experiment, the parts are assembled and the flexure frame is manually compressed. The clip secures the gap height prior to loading the sample. By releasing the clip abruptly, a step strain is generated and the liquid filament deforms. (b) Here we show an example of an artifact in the step strain on PocketCaBER, demonstrated by the distance between plates $d$ as a function of time $t$. Depending on the stiffness of the flexure frame material, early-time oscillations are observed around the the step-strain function of gap height versus time. These oscillations can be tuned/minimized by varying the printing material, as shown in this plot. (c) Here we show snapshots of an experiment using PocketCaBER on a PIB solution (approximately \qty{1e6}{g/mol}, $\sim0.3$wt\% in mineral oil). After the plates separated, a filament formed and gradually thinned until it broke up. Filament vibration was observed at early times (\textit{e.g.}, $t=\qty{0.04}{s}$) due to inertial effects. Images are obtained from a 4K 120fps slow-motion video taken using iPhone 17 Pro, with an attached commercial macro lens (SmallRig \qty{75}{mm}).}
    \label{Fig:1}
\end{figure*}

\section{PocketCaBER: A 25-cent Capillary Breakup Extensional Rheometer}

\begin{figure*}[ht!]
    \centering
    \includegraphics[width=\textwidth]{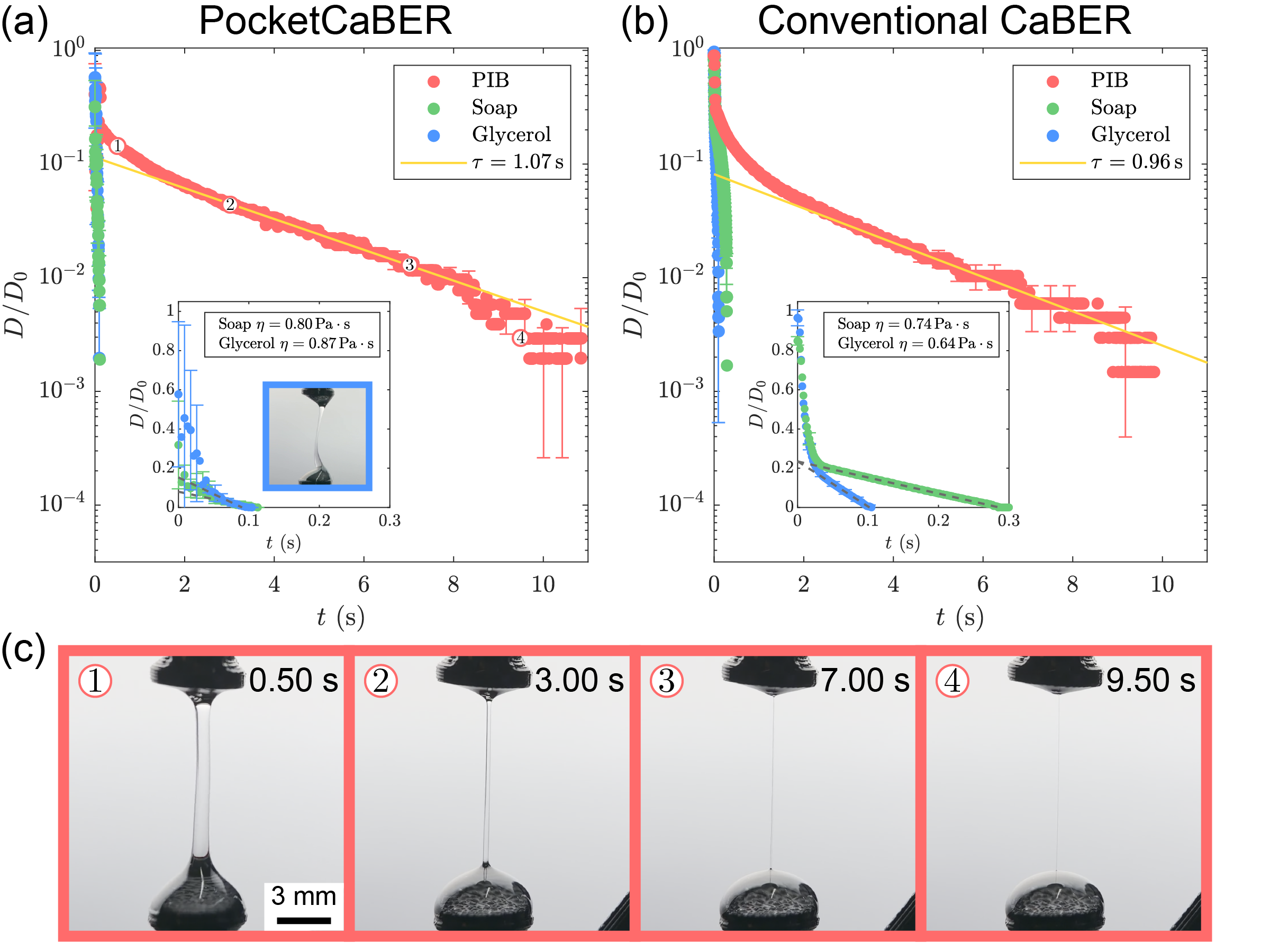}
    \caption{\textbf{Comparison between measurements from PocketCaBER and conventional CaBER} (a) Here we show the results from PocketCaBER for three samples: PIB solution (same as Fig.\ref{Fig:1}(c)) (red), Dawn dish soap (green), and glycerol (blue). All data were extracted from videos of experiments with a custom MATLAB script. PocketCaBER data were acquired with an iPhone 17 Pro using slow-motion mode, with frame rates of 120fps (PIB), 240fps (soap), and 240fps (glycerol). To fit an extensional relaxation time $\tau$ to the PIB data, a window between $t=\qty{2}{s}$ and $t=\qty{6}{s}$ was used, yielding $\tau=\qty{1.07}{s}$.  (b) We show the same three fluids on a traditional CaBER (device located at MIT). The videos were taken with a Phantom high-speed camera (VEO model), with frame rates of 120fps (PIB), 600fps (soap), and 600fps (glycerol). For relaxation time of PIB, a window between $t=\qty{4}{s}$ and $t=\qty{8}{s}$ was used, yielding $\tau=\qty{0.96}{s}$. Additionally, viscosity was fitted for soap and glycerol, yielding values shown in the respective inset of the plots. An example experimental snapshot of glycerol (using PocketCaBER) is shown in the inset of (a). (c) Four snapshots from an experiment video of PocketCaBER on the PIB solution are shown, indicating the filament thickness in three different regimes of filament thinning. Figure \textcircled{1} ($t=\qty{0.50}{s}$) corresponds to the visco-capillary (VC) regime; figures \textcircled{2} ($t=\qty{3.00}{s}$) and \textcircled{3} ($t=\qty{7.00}{s}$) corresponds to the elasto-capillary (EC) regime; figure \textcircled{4} ($t=\qty{9.50}{s}$) corresponds to the finite extensibility regime. }
    \label{Fig:2}
\end{figure*}

\subsection{PocketCaBER Design}

Conventional CaBER equipment leverages a linear precision stage to achieve the extensional step-strain that leads to the sequential capillary thinning of the sample filament. In order to accomplish this step-strain, we chose to use a 3D-printable flexure mechanism for PocketCaBER to replace the traditional linear stage. The thickness of the PocketCaBER is relatively flat (\qty{3}{mm} thick) with a width of about \qty{100}{mm} and a height of roughly \qty{60}{mm}. It can be printed in roughly 30 minutes on a hobby 3D printer due to design compatibility with print-bed orientation. The device can be subsequently assembled via either press fit or reinforced with adhesive or epoxy. The assembled device is highly portable and does not require any electrical microcontrollers or motors. Because of the design's simplicity, it is well-suited to expand pedagogical capability in larger classroom environments, where access to power, access to suitable instrument controllers, or restrictions on the individual participant's physical space has traditionally constrained curriculum to lecture-based coursework. Moreover, the design is also suited to expanding the field of extensional rheology to field environments, in regions of the world where cold-chain and refrigeration are challenging---and scenarios where the material mutates, such that minimizing sample transit time is critical.
As described in \citet{collett2015portable}, others have attempted to build traditional rheological field tools in this spirit. This tool is designed to facilitate research in these scenarios where few tools exists that are both sufficiently accurate, practical and low cost.

For data analysis with this method, a cell phone camera that is able to take slow-motion videos (120 frames-per-second (fps) or 240fps in this study) replaces the traditionally-used high-speed camera in CaBER, making it more accessible and facilitating ease-of-use. The whole setup is shown in Fig. \ref{Fig:0}(b). Modification can be made to the setup to include a cell-phone macro lens and back-lighting to achieve better video and data quality.

The procedure for using PocketCaBER to make measurements is relatively simple compared to conventional CaBER. The standard procedure is shown in Fig. \ref{Fig:1}(a). To operate the device, the plates are positioned parallel to each other. Before starting an experiment, the flexure frame is compressed and the clip is affixed onto the ridges along the frame. The distance between the sample plates can be manually adjusted to be roughly the same as the plate radius (\qty{3}{mm} in the present design). With the sample loaded between the plates, the camera is aligned with the test region, preferably evenly lit with a solid background. We start recording in slow-motion mode (120fps or higher), and subsequently flick the clip off the flexure (and away from the sample region) to release the flexure and initiate an extensional step-strain on the sample. This release starts the capillary thinning experiment.

It is notable that different polymeric feedstock (for common hobby 3D filament printers) leads to variable device performance. Particularly, the stiffness and springiness of the device depends on the chosen material, which further results in different oscillation behavior of the liquid filament in the IC/VC regime. To demonstrate this effect, we tested three materials: carbon fiber enhanced polyethylene terephthalate glycol (PETG-CF), carbon fiber enhanced polylactic acid (PLA-CF) and polylactic acid (PLA). In order to test the response of the device, we plotted the distance between plates as a function of time from flexure release, as shown in Fig. \ref{Fig:1}(b).  The opening time of the flexure is approximately \qty{0.01}{s}, which is reasonable to be considered as a step-strain. As shown, PETG-CF has stronger under-damped oscillatory behavior when compared with both PLA and PLA-CF. This early-time oscillation (in all cases under 30--40 milliseconds) leads to an artifact in the step strain, which then can propagate unwanted effects to the fluid---as shown in the third panel of Fig. \ref{Fig:1}(c). This can limit measurements in the IC/VC regime. While these limitations restrict sample resolution for Newtonian-like fluids, we observe that these under-damped oscillations at short times do not appear to significantly impact the EC regime for samples with longer relaxation times (for example, fluids with relaxation times on the order of 1 second appear unaffected in the EC regime). The whole capillary thinning process of a Boger fluid ($\sim 0.3$wt\% with $\approx 10^6\,\mathrm{g/mol}$ polyisobutylene (PIB) in mineral oil) upon separation of the plates is shown in Fig. \ref{Fig:1}(c). By tracking the diameter of the formed liquid filament, it is possible to extract multiple material properties of the sample, including the extensional relaxation time $\tau$ and the time to break up $t_b$.

\subsection{Proof of Concept and Comparison to State-of-the-Art}

To evaluate PocketCaBER performance, we conducted experiments on three different samples: the PIB Boger fluid mentioned previously, dish soap (Dawn) and glycerol. We then compared the results with the gold-standard CaBER (located at MIT) using the same three samples, shown in Fig. \ref{Fig:2}(a--b). For each sample, the experiments were repeated three times (with the same initial geometry, \textit{i.e.}, gap distance between plates). The experiments for PocketCaBER were recorded with an iPhone 17 Pro using slow-motion mode, with frame rates of 120fps for PIB solution and 240fps for soap and glycerol. The experiments for standard CaBER were taken with a Phantom high-speed camera (Vision Research, VEO model), with frame rates of 120fps for PIB solution and 600fps for soap and glycerol. All the videos were then processed and analyzed with a custom MATLAB script, extracting normalized filament diameter $D/D_0$ as a function of time $t$, where $D_0=\qty{6}{mm}$ is the plate diameter. For the non-Newtonian PIB Boger fluid, an extensional relaxation time was inferred using Eq. \eqref{radius} in the EC regime. For soap and glycerol, a rough estimate of their respective viscosities was inferred from the expression given by \citet{Mckinley2000}:
\begin{equation}
    \frac{D}{D_0}=\frac{(2X-1)}{3}\frac{\Gamma}{\eta D_0}(t_b-t),\quad X=0.7127,
    \label{viscosity}
\end{equation}
where $X$ is constant empirical parameter, shown to equal approximately 0.71 for a Newtonian fluid \citep{Papageorgiou1995}. The surface tension $\Gamma$ was measured by a DCAT25 DataPhysics Tensiometer, as given by the values in Table \ref{tab:1}. It is notable in our experiments that frames close to filament breakup are quite sensitive to image binarization and thresholding parameters in the MATLAB script. If using PocketCaBER for quantitative analysis near filament breakup, we note that the most sensitive image thresholding parameters are flagged in the analysis code, and do require fine-tuning for best results.

The PIB Boger fluid exhibited three distinct filament thinning regimes (VC, EC and finite extensibility), with each clearly distinguishable in cell phone video footage with a macro lens. As demonstrated in Fig. \ref{Fig:2}(c), experimental frames are shown for each of the corresponding regimes of filament thinning. For the extracted extensional relaxation times, PocketCaBER gives a result of $\tau=\qty{1.07}{s}$ and standard CaBER gives $\tau=\qty{0.96}{s}$, which are consistent to within $\sim 10$\%. Note that the choice of onset and end times of the EC regime for fitting can influence the result of the relaxation time by roughly this same order of magnitude ($\sim 10$\%). The results presented here suggest that PocketCaBER is capable of measuring model-specific rheological properties---\textit{i.e.}, extensional relaxation time---to a remarkably similar level of accuracy as research-grade gold-standard equipment. Additionally, the sizes of the error bars are reasonably small to indicate the reproducibility of the experiments on PIB solution, further demonstrating the performance of PocketCaBER on an example non-Newtonian fluid with a relatively long relaxation time ($\sim\qty{1}{s}$).

However, for Newtonian liquids (dish soap and glycerol, shown in the inset of Fig. \ref{Fig:2}(a)) in our experiments, the results are less ideal due to the aforementioned vibration-induced inertial effects, and oscillation of the liquid filament. Using Eq. \eqref{viscosity}, we measured dish soap and glycerol viscosities of 0.80 and \qty{0.87}{Pa\cdot s}, respectively. Compared with the rheometer-based viscosity measurements shown in Table \ref{tab:1} (0.98 and \qty{0.79}{Pa\cdot s} for dish soap and glycerol), this technique is reasonable in some cases to estimate a rough order of magnitude. However it is noted the error was significant between Newtonian fluid trials ({\textit{i.e.}, the standard deviation of $D/D_0$  exceeded 0.2 in some early frames of both Newtonian fluids), and caution should be taken when attempting to characterize these fluids. It is also noted that the fitted viscosity values from standard CaBER (0.74 and \qty{0.64}{Pa\cdot s} for dish soap and glycerol) deviate from rheometer measurements as well, reflecting the limitations of CaBER on Newtonian fluids. We conclude that PocketCaBER is well-suited (and indeed comparable to state-of-the-art) when measuring viscoelastic fluids in the EC regime with relaxation times on the order of \qty{1}{s}, but exhibits significant error when measuring IC/VC properties of more Newtonian fluids (with very short relaxation times $\ll 0.1\,\mathrm{s}$) due to this specific flexure-based design. 

\begin{figure*}[ht!]
    \centering
    \includegraphics[width=\textwidth]{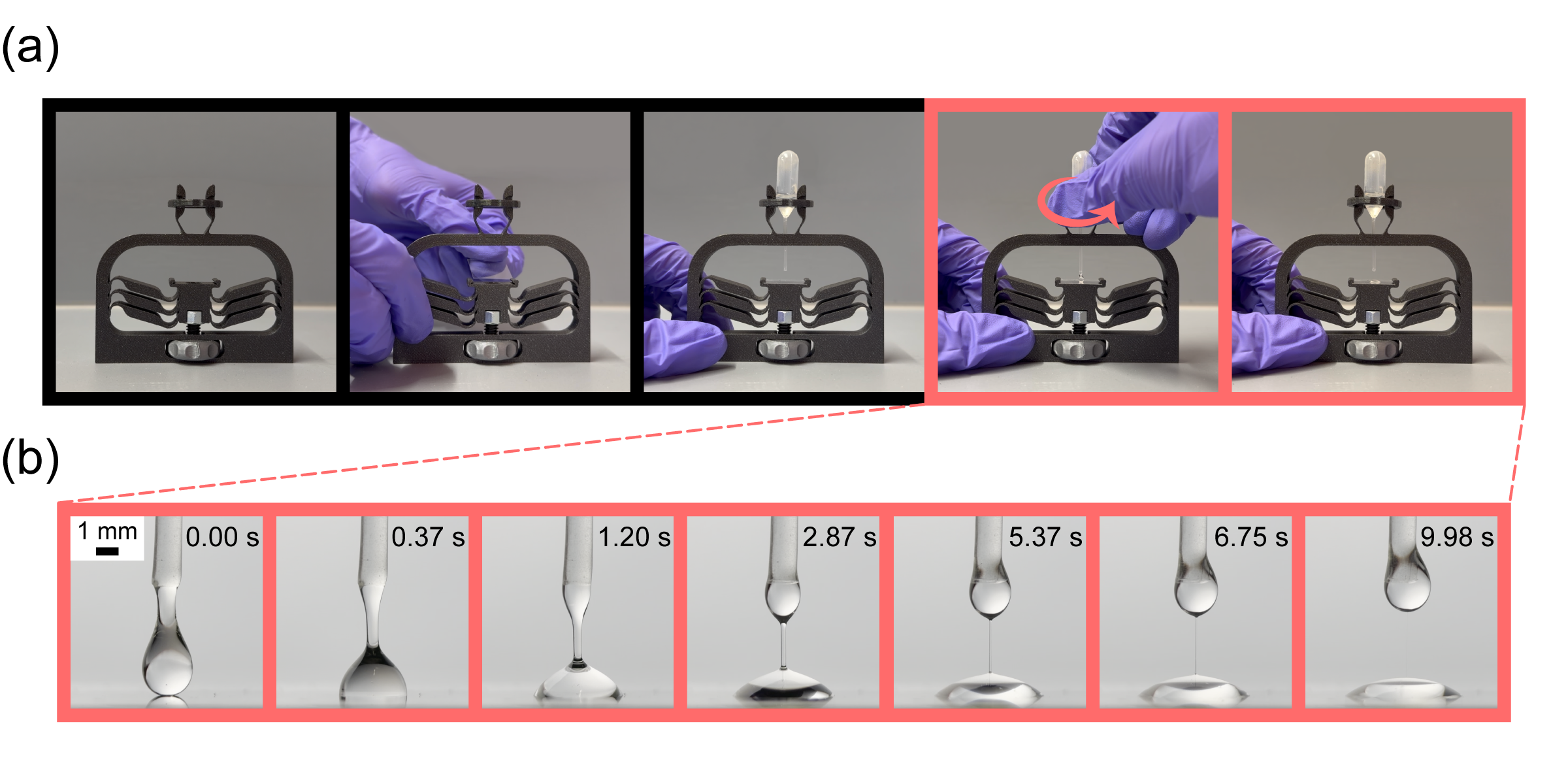}
    \caption{\textbf{Design of PocketDoS} (a) The experimental protocol of PocketDoS is shown. To set up an experiment, the parts are assembeld and and the cam is placed around the pipette holder. A coverslip is then loaded onto the flexure stage. A sample is then loaded into a small plastic transfer pipette and placed in the top pipette holder. The cam can be rotated to compress the transfer pipette and apply a small pressure, to facilitate the formation of a droplet. This droplet may either be gently dripped onto the coverslip, or the stage may be moved up by rotating the lower adjustment knob (which lifts the flexure stage smoothly by interfacing with a 1/4''-20 bolt, which acts as the lead-screw in this design). Once the droplet attaches to the lower surface, the liquid filament evolves and thins until it breaks. (b) Snapshots of an experiment using PocketDoS on a PIB solution (same as Fig.\ref{Fig:1}(c)) are shown. After the droplet attached to the substrate (coverslip), a filament of PIB solution formed and gradually thinned until it broke up. Images are obtained from an HD 120fps slow-motion video taken using iPhone 17 Pro, with a SmallRig \qty{75}{mm} macro lens.}
    \label{Fig:3}
\end{figure*}

\section{PocketDoS: An Ultra-Frugal Approach to Dripping-onto-Substrate (DoS) Extensional Rheometery}

\begin{figure*}[ht!]
    \centering
    \includegraphics[width=\textwidth]{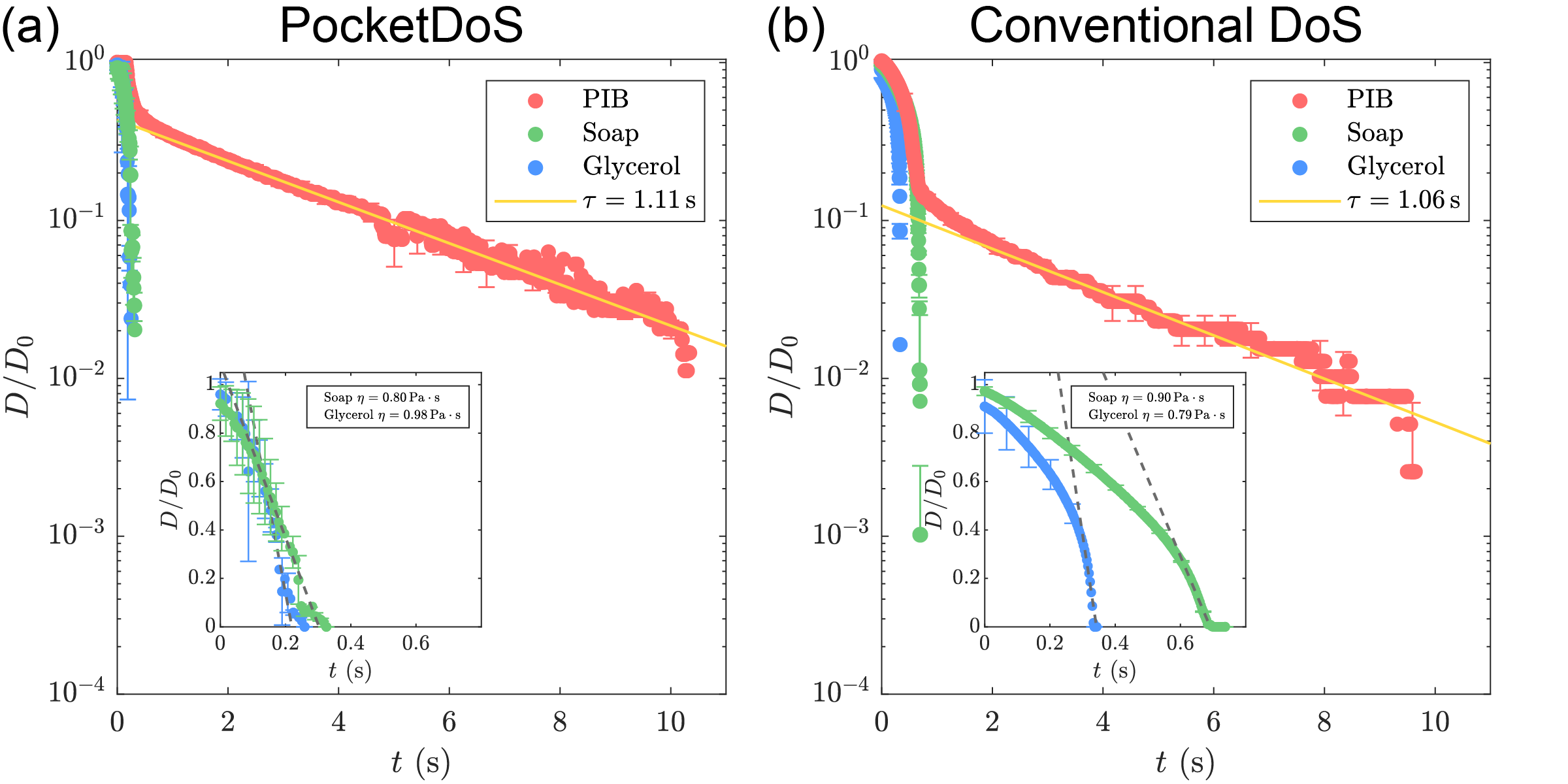}
    \caption{\textbf{Comparison between measurements from PocketDoS and conventional DoS} (a) Shown here are the results from PocketDoS for three samples: PIB solution (same as Fig.\ref{Fig:1}(c)) (red), Dawn dish soap (green), and glycerol (blue). All data were extracted from videos of experiments with a custom MATLAB script. PocketDoS data were acquired with an iPhone 17 Pro using slow-motion mode, with frame rates of 120fps (PIB), 120fps (soap), and 120fps (glycerol). To fit an extensional relaxation time $\tau$ to the PIB data, a window between $t=\qty{2}{s}$ and $t=\qty{4}{s}$ was used, yielding $\tau=\qty{1.11}{s}$. (b) We shown the same three fluids on a traditional DoS (device located at MIT). The videos were taken with a Phantom high-speed camera (VEO model), with frame rates of 120fps (PIB), 300fps (soap), and 300fps (glycerol). For relaxation time of PIB, a window between $t=\qty{2.3}{s}$ and $t=\qty{6.3}{s}$ was used, yielding $\tau=\qty{1.06}{s}$. Additionally, viscosity was fitted for soap and glycerol, yielding values shown in the respective inset of the plots.}
    \label{Fig:4}
\end{figure*}

\subsection{PocketDoS Design}

Regular DoS setup involves a syringe pump that dispenses a droplet which subsequently forms a thinning filament upon contact with the substrate. We designed a rotating cam with a pincer for a small plastic transfer pipette to replace a standard laboratory syringe pump. The substrate is a thin standard glass coverslip. We leveraged the same flexure mechanism design for the sample holder stage so that the substrate can be moved slowly towards the droplet for contact with a rotating knob (lead screw is replaced with a standard $1/4$''-$20$ bolt). Similar to PocketCaBER, our design for PocketDoS is small and easily printed on a hobby 3D printer. A cell phone camera is used for capturing slow-motion videos for data analysis; an off-the-shelf cell-phone macro lens attachment is more necessary in this setup because the filament diameters are smaller, requiring more magnification. The experimental setup is shown in Fig. \ref{Fig:0}(d).

To assemble the device, the cam is placed around the top pipette holder, the bolt, stage adjuster and nut are assembled under the flexure and a coverslip is positioned on top of the precision-adjustable flexure sample stage. The standard operating procedure for PocketDoS, shown in Fig. \ref{Fig:3}(a), involves loading a sample into a disposable plastic transfer pipette and placing it within the top cam. The pipette is softly deformed when the cam is rotated, qualitatively setting pressure at the nozzle. The cell phone camera is then aligned with the test region with an evenly illuminated solid background. We begin recording in slow-motion mode (120fps or higher), and subsequently rotate the cam to dispense a droplet from the tip of the pipette nozzle. To initiate the capillary thinning process, a droplet is then slowly deposited onto the coverslip by moving the stage up slowly or by gently dripping from a fixed height above the coverslip (where $h_{\text{drop}} \sim 2R/3$, with the bottom of the droplet at $t=0$ as $h_{\text{drop}}$ and the radius of the droplet as $R$). There are pros and cons to both approaches. Direct deposition of the droplet is easier to operate, but the initial flow from the nozzle leads to moving contact lines, which influences the capillary thinning process and consequentially the measured material properties. Moving the stage up to the droplet mitigates these issues, but is a more challenging technique for an individual operator.

The capillary thinning process of a slender filament of the PIB solution is shown in Fig. \ref{Fig:3}(b). Similar to PocketCaBER, by tracking the diameter of the liquid filament, rheological properties of the sample can be inferred. However, unlike PocketCaBER, the formation of the liquid filament is less influenced by oscillation of the device; thus PocketDoS is better suited to low-viscosity sample measurements.

\subsection{Proof of Concept and Comparison to State-of-the-Art}

For evaluating PocketDoS performance, we also conducted experiments on the same three samples as PocketCaBER: the PIB Boger fluid, Dawn dish soap and glycerol. Additionally, we compared the results with the research-grade DoS (located at MIT) using the same three samples. The experiments for each sample are shown in Fig. \ref{Fig:4}, with data from multiple trials per solution. The experiments for PocketDoS were recorded with an iPhone 17 Pro using slow-motion mode, with frame rates of 120fps for all three liquids. The experiments for standard DoS were taken with the aforementioned Vision Research high-speed camera, with frames rates of 120fps for PIB solution and 300fps for soap and glycerol. All videos were then processed and analyzed with the same image processing library as PocketCaBER (see: Data Availability section). The normalized filament diameter $D/D_0$ was then extracted as a function of time $t$ ($D_0=\qty{1.27}{mm}$ was the nozzle diameter in these experiments). For the non-Newtonian PIB solution, an extensional relaxation time was inferred using Eq. \eqref{radius} in the EC regime. For soap and glycerol, a rough estimate of the respective viscosities was inferred from \eqref{viscosity}.

For PIB Boger fluid, the extracted extensional relaxation time with PocketDoS was $\tau=\qty{1.11}{s}$, and the same quantity obtained with research-grade DoS was $\tau=\qty{1.06}{s}$; within 5\% deviation from each other. The size of error bars in Fig. \ref{Fig:4}(a) for PIB are comparable to the error bars in Fig. \ref{Fig:4}(b)---indicating the reproducibility of PocketDos is remarkably similar to lab-grade research DoS setups for viscoelastic fluids. This demonstrates PocketDoS is capable of extracting model-specific relaxation times for fluid samples with relaxation times on the order of $\sim\qty{1}{s}$, in a manner similar to benchtop DoS setups.

However, when assessing the performance of PocketDoS with Newtonian liquids (dish soap and glycerol), the error is more significant in comparison to benchtop DoS setups (Fig. \ref{Fig:4}(a). Using Eq. \eqref{viscosity} and measured surface tension shown in Table \ref{tab:1}, we obtained dish soap and glycerol viscosity values of 0.80 and \qty{0.98}{Pa\cdot s}, respectively. These results are reasonably close to measurements made on the rheometer (shown in Table \ref{tab:1}; 0.98 and \qty{0.79}{Pa\cdot s} for dish soap and glycerol). This suggests that PocketDoS may be suitable in some cases to estimate a rough order of magnitude for sample viscosity. However, camera sensor and illumination limitations restrict its performance relative to traditional setups (as shown by the relative size of the error bars in the inset panels of Fig. \ref{Fig:4}(a) versus Fig. \ref{Fig:4}(b)). Indeed, the fitted viscosity values (same linear regression method) from standard DoS were 0.90 and \qty{0.79}{Pa\cdot s} for dish soap and glycerol---which are significantly closer to the ``ground truth'' rheometer values, demonstrating a better performance of traditional benchtop DoS on low-viscosity Newtonian fluids. We note that the values here were fitted with data close to filament breakup to avoid inertia effects. In conclusion, PocketDoS is an appropriate technique to measure viscoelastic fluids in the EC regime with relaxation times on the order of \qty{1}{s}, but is less ideal for measurement of viscosity of more Newtonian fluids.

\section{Discussion}
We have shown that for Newtonian fluids and example viscoelastic fluids, the device designs are approximately equivalent with state-of-the-art laboratory capillary breakup and dripping-onto-substrate methods. Our pocket devices are capable of giving rough estimated values of viscosity for Newtonian fluids. More importantly, they excel at extracting the extensional relaxation time for viscoelastic fluids similar to or greater than $\sim1\,\mathrm{s}$, within $\sim10\%$ deviation from results obtained from research-grade gold-standard equipment. Additionally, the fitted relaxation times for the samples given by PocketCaBER and PocketDoS were comparable, further underscoring both devices' repeatability and reliability. 

\subsection{Protorheology vs. Model Agnostic vs. Model Specific}
These devices in teaching contexts can be used in various different ways to measure the properties of fluids. Here we delineate between qualitative assessments of fluids (``protorheology'', \textit{cf.} \citet{Hossain2024}) where sequential tests may be used to determine if one solution is more elastic than another---or if a solution is Newtonian vs. viscoelastic in character. These ``proto'' tests typically result in yes/no or ``sample A more than sample B'' type results, and in teaching contexts and educational outreach can prove more accessible and an easy starting point to conversations about extensional flows and advanced fluid mechanics. Protorheology can also inform some quantitative properties such as time-to-breakup measurements for more sophisticated comparisons between complex fluids, as shown in table \ref{tab:discussion}. 

One step more refined in characterization would involve model-agnostic quantitative material parameters such as the extensional viscosity versus time, which can be inferred from image processing data and a known surface tension measurement. The extensional viscosity function is summarized for Newtonian versus non-Newtonian fluids in \ref{tab:discussion}. This describes properties that are representative of the bulk that may evolve in time in the sample---for example extensional thickening at high strain amplitudes in an extensional flow. This data allows quantitative comparisons between samples for time-dependent or transient material properties that physically describe the behavior of the fluid in response to a strain history. 

Finally, it is possible to extract model-specific parameters that are predictive of the relationship between stress and strain (and strain rate) for a given sample. These model parameters use the experimental data to fit to a generalized constitutive model---such as a single-mode upper-convected Maxwell model or Oldroyd-B model. Relaxation time is an example of a model-specific parameter, which, when used in conjunction with a specific model, can describe that fluid's behavior generally across different strain histories and flow types. The drawback to performing model-specific analyses is that there are underlying assumptions that the model choice is appropriate to the fluid sample, and the fitted parameters are representative of the behavior in regimes that may require further data (\textit{i.e.}, projecting the model out into regimes that may be not valid). However, these model-specific parameters are a powerful low-dimensional quantification of material behavior. For example, relaxation time (in a teaching context) enables users to quantify ``how viscoelastic'' is a fluid, as quantified by a single number---which is a useful method to rigorously contrast materials.


\begin{table*}[htbp]
    \centering
    \caption{Comparison of properties measured at different levels.}
    \label{tab:discussion}
    \renewcommand{\arraystretch}{1.25}
    \setlength{\tabcolsep}{4pt}
    \begin{tabularx}{\textwidth}{@{}>{\raggedright\arraybackslash}p{0.12\textwidth}>{\raggedright\arraybackslash}X>{\raggedright\arraybackslash}X>{\raggedright\arraybackslash}X>{\raggedright\arraybackslash}p{0.35\textwidth}@{}}
    \toprule
    Level & Measured quantities & Assumptions & Material functional property & Equation \\
    \midrule
    Proto-rheology & Time & --- & Time to breakup, $t_b\,[\mathrm{s}]$ & --- \\
    Model agnostic & Radius $R$, surface tension $\Gamma$ & --- & Newtonian shear viscosity, $\eta\,[\mathrm{Pa\,s}]$ & {\small$\displaystyle\frac{R}{R_0}=\frac{(2X-1)}{6}\frac{\Gamma}{\eta R_0}(t_b-t),\quad 0.5324 \leq X \leq 1$} \\
    & & & Newtonian extensional viscosity, $\eta_E\,[\mathrm{Pa\,s}]$ & {\small$\displaystyle\eta_E=3\eta$} \\
    & & & Non-Newtonian (transient) extensional viscosity, $\eta_E^+\,[\mathrm{Pa\,s}]$ & {\small$\displaystyle\eta_E^+(t)=-\frac{1}{2}\frac{\Gamma}{\mathrm{d}R/\mathrm{d}t}$} \\
    Model specific & Radius $R$ & UCM/Oldroyd-B & Relaxation time, $\tau\,[\mathrm{s}]$ & {\small$\displaystyle\frac{R}{R_0}=\left(\frac{G R_0}{2\Gamma}\right)^{1/3}\exp\!\left[-\frac{(t-t^\ast)}{3\tau}\right],\quad t \geq t^\ast$} \\
    \bottomrule
    \end{tabularx}
\end{table*}

\section{Teaching Advanced Fluid Mechanics Topics via Experiential Learning}

\begin{figure*}[ht!]
    \centering
    \includegraphics[width=1\linewidth]{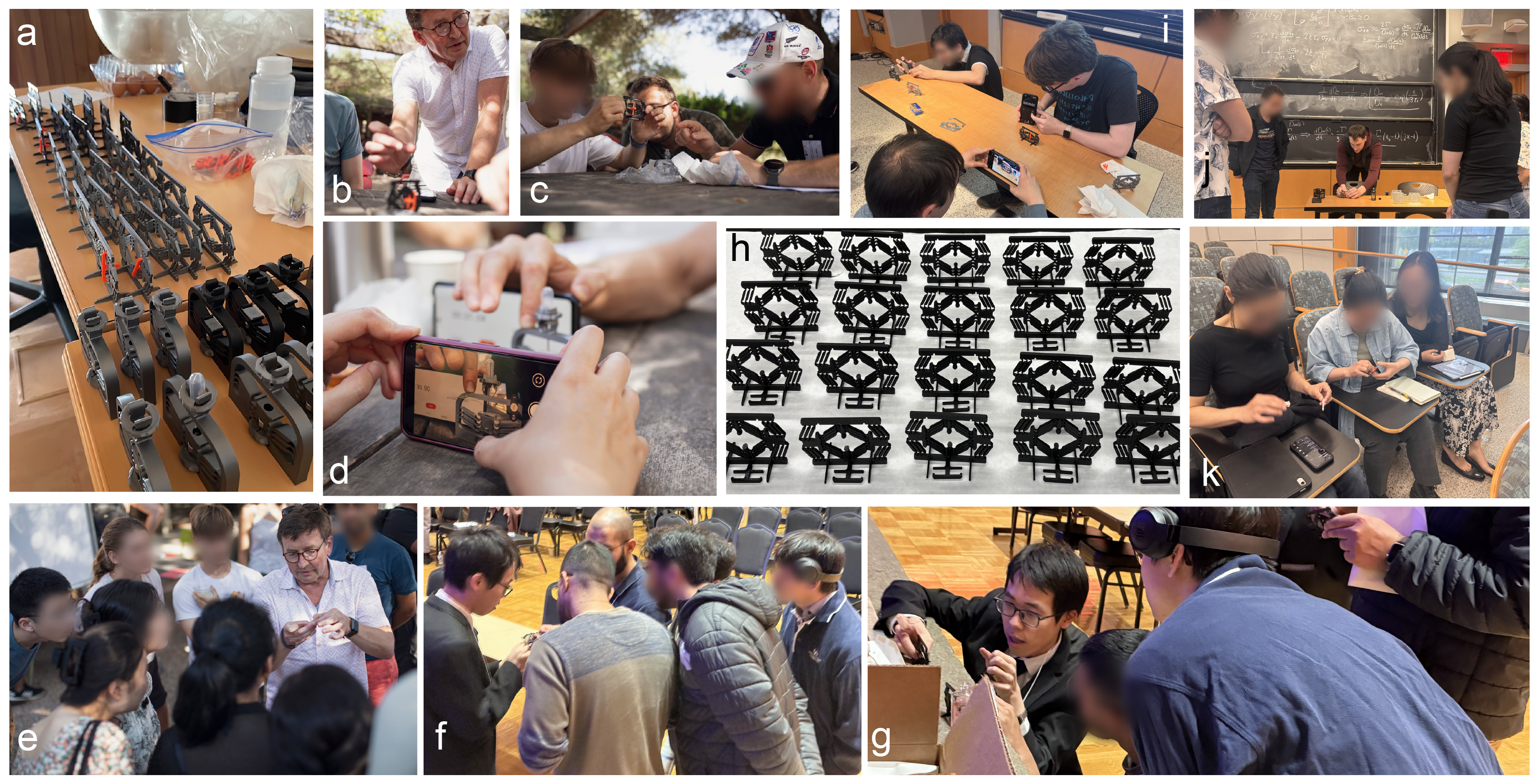}
    \caption{\textbf{Experiential learning in action} (a-e) PocketCaBER and PocketDoS Extensional Rheology Mini-Lab at the L'Institut d'Études Scientifiques de Cargèse summer school on soft matter in Corsica (Kroo/McKinley teaching; photo credit 6b-e to Thibaut Divoux), (f-h) Spontaneous PocketCaBER tutorial and mini-lab shortly after lightning talk by Zhaofeng Peng at UMass Amherst's inaugural College of Natural Sciences Graduate Student Research Symposium, (i-k) Warwaruk/McKinley teaching extensional rheology at MIT in 2.341J Spring 2025.}
    \label{fig:teaching}
\end{figure*}

PocketCaBER and PocketDoS devices have now been incorporated into advanced graduate-level curriculum in multiple programs in the United States, including: Massachusetts Institute of Technology (course 2.341J), University of Massachusetts Amherst (Polymer 704) and University of Illinois Urbana-Champaign (TAM 534/498). Internationally, these devices were also employed at a 2024 summer school on soft matter held at l'institut des Études Scientifiques (IES) in Cargèse, Corsica. From these numerous and diverse teaching contexts, we observed several overarching pedagogical insights, based on class feedback/findings and testing outcomes when using these devices to supplement traditional methods for teaching and learning advanced topics in fluid mechanics.   

 One insight was that extreme cost-minimization of this instrument design enables feasibility of rolling out these tools at scale for lab modules involving large class sizes. For example, the Cargèse summer school had over 70 participants and an additional 10--15 lecturers (nearly all of whom participated in these labs). Because our design does not require motorized stages, linear bearings or microcontrollers, these flexure-based instruments cost roughly 25 cents per device and can be manufactured in bulk. Depending on the chosen material and print settings, each device (PocketCaBER specifically) requires roughly 10--\qty{11}{g} of filament; equating to roughly 25 cents for PLA (\qty{10.78}{g} at $\approx 23$ USD per kilogram) or 36 cents for PETG (\qty{10.37}{g} at $\approx 33$ USD per kilogram), based on the cost of hobby 3D printer filament (for a Bambu labs H2D pro printer) at the time of publication. Ordering in bulk at scale would even further reduce this per-device cost. The benefit of this cost-minimzation went beyond scalability in these teaching environments---from a practical perspective, it meant that many participants could retain the devices for their own personal use after the program. This noticeably increased engagement: when tools are \textit{distributable} during a class such that the lab bridges traditional teaching environments and research environments, we observed participants engaging with the subject matter differently.   
 
 This observation about accessibility of instrumentation led our team to another central insight: the success of these teaching tools is not solely about cost-minimization: the \textit{simplicity} of design and user-oriented operation allowed for  participants to focus on fluid measurements rather than debugging software or rigorously sweeping through a parameter space. This minimal simplicity in the teaching environment enables focus on the physics of fluids---rather than details around engineering and instrument setup. 

\subsection{Rheology Tools Online Repository}
With the wide spread accessibility of 3D printers, several large ``DIY'' communities have flourished online as sources for collaboration, discussion, and iteration onto home made and custom scientific devices. Recently, the field of rheology has begun to build an online community platform on GitHub to share resources for the development, construction, and implementation of DIY rheology tools. The GitHub could potentially provide access to instruction manuals, CAD design files, and programming code that enables the immediate rheology community to conduct their own rheology experiments. Establishing a central online presence is critical for the eventual goal of tapping into the existing ``DIY'' communities and sharing the science of rheology with the broader public community. The ``DIYRheoHub'', launched in March 2026, is a central landing spot hosted through GitHub Pages (url---https://rheohiolab.github.io/DIYRheoHub.github.io/) that highlights several published DIY rheometer designs. Projects are defined by their cost, difficulty of construction, and required materials (3D printer, electronics, \textit{etc.}) to help users decide which device will fit their means and abilities. Vistors to the website will find links to other papers, repos, or personal websites that contain the required details for construction. As a flexible living document, the DIYRheoHub is constantly updated by the curators. Members of the user community are also able to contribute their own content by pushing to the repo, pending approval and modification from the curators. The scope of the ``DIYRheoHub'' is to promote open source accessibility. Therefore, all projects are properly cited, but also abide by the copyright privileges associated with the source material. A future goal is to consolidate instruction manuals and programming codes into dedicated GitHub repos to promote collaboration within a single online platform.

\section{Conclusions}
Open-source instrumentation such as PocketCaBER and PocketDoS open opportunities for experiential learning and field-science that were not previously accessible. By using principles of rapid prototyping and frugal design (such as flexure mechanisms and cell phone camera optics), we are able to extend the field of frugal science to extensional rheometry; enabling advanced teaching and learning both within the classroom and in field-research settings. PocketCaBER and PocketDoS are further examples of the rheological communities recent push to provide open access through GitHub resources online. The growing number of devices will be shared, iterated, and advanced upon in the common pursuit to share rheology principles and characterization with the general public.
  

\section*{Appendix}
\appendix

\begin{figure}[ht]
    \centering
    \includegraphics[width=1\linewidth]{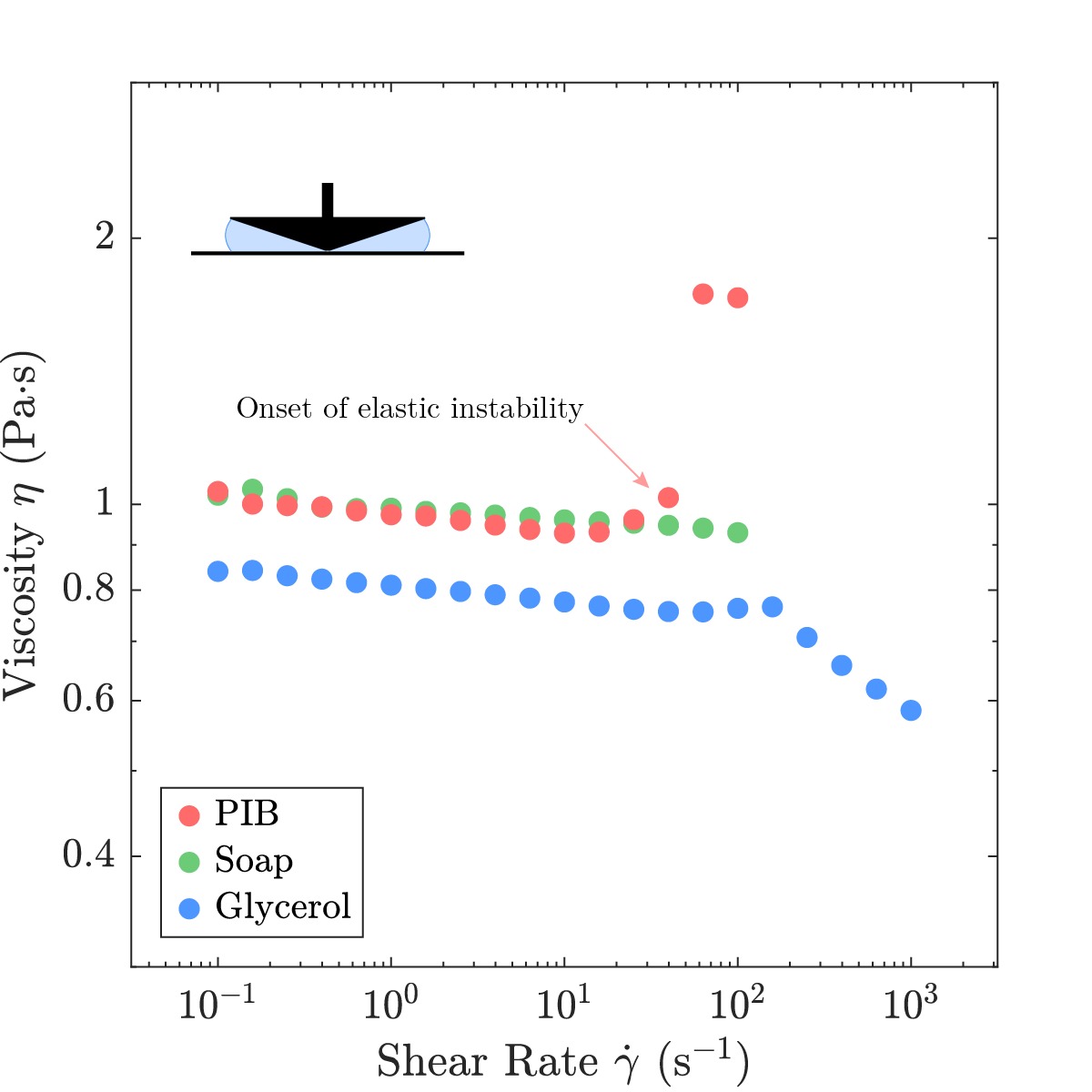}
    \caption{\textbf{Steady state flow curves} The viscosities of the three samples in this study were measured using a \qty{60}{mm} $2^\circ$ cone-plate geometry at \qty{25}{\degreeCelsius} on a TA Instruments Discovery HR rheometer. The viscosities used in this paper were taken from these measurements in the linear regime shown (under \qty{100}{s^{-1}}). Above this there are some artifacts in the DHR measurements from torque limits and elastics.}
    \label{fig:viscosity}
\end{figure}

\begin{table}[ht]
	\caption{Material properties of the different test fluids.}\label{tab:1}%
	\begin{tabular}{@{}cccc@{}}
		\toprule
		Fluid & $\rho$ (kg m$^{-3}$) & $\eta$ (Pa s) & $\Gamma$ (mN m$^{-1}$)\\ 
		\midrule
		Dish soap & $1007\pm14$ & $0.98\pm0.06$ & $25.8\pm0.9$ \\
		Glycerol & $1251\pm13$ & $0.79\pm0.06$ & $62.8\pm0.2$ \\
		PIB Boger fluid & $840\pm8$ & $0.97 \pm 0.07$ & $30.7\pm0.2$ \\
		\bottomrule
	\end{tabular}
\end{table}

The rheology of the test fluids used in this study was measured on a controlled stress rheometer (Discovery HR, TA Instruments, New Castle, DE) using a \qty{60}{mm} cone-plate configuration at \qty{25}{\degreeCelsius}. The surface tension was measured using a du No\"uy ring configuration (DCAT Dynamic Tensiometer). Results are tabulated in Table \ref{tab:1} and steady state flow curves are given in Figure \ref{fig:viscosity}. The sudden upturn in the data for the PIB solution corresponds to onset of an elastic instability \citep{more2024elasto}.

\subsection*{Video Analysis Protocol for Pocket Devices}
\begin{enumerate}
    \item[1)] The slow-motion videos recorded with the cell phone are first exported to a computer. Then, the videos are converted to \texttt{*.tif} images for all the frames, with software such as FFmpeg. The user can then decide the region in each frame containing the liquid filament to be analyzed and the plate diameter (for PocketCaBER)/the nozzle width (for PocketDoS) in pixels with ImageJ.
    \item[2)] With all the frame images stored in the same folder, they are now ready to be analyzed with the MATLAB script. To start with, the user should specify the parameters obtained from the first step (analysis region, frame rate of the video in fps and geometry parameters). A fine-tune of the image binarization threshold parameter is required if the illumination condition/video quality is less ideal (which will result in a blurry filament boundary, \textit{etc.}).
    \item[3)] Using the MATLAB script, the user can plot the data to check the quality of the analysis. By specifying the lower and upper bounds (of time) of the elasto-capillary, a fit for the relaxation time can be obtained.
\end{enumerate}

\section*{Acknowledgments}
We would like to acknowledge the many students and summer school participants who helped the authors adapt and develop these instruments for use in the classroom. Additionally, we would like to thank Thibaut Divoux for organizing (and photographing) the Cargèse summer school. 

\section*{Author Contributions}
LK and GHM conceptualized the project. LK and ZP designed and printed the PocketCaBER and PocketDoS designs. ZP and LW collected all data. ZP analyzed all data, wrote manuscript and made figures. ZP, LK, LW BY, TL, GHM and RE editted manuscript. ZP, LK, LW, TL, RE, GHM taught courses using the devices for insights on teaching sections. 


\section*{Data Availability}
The data are available from the corresponding author upon reasonable request and will be publicly available online at time of peer reviewed manuscript acceptance.

\section*{Declarations}
\subsection*{Conflict of Interest}
The authors declare no competing interests.

\bibliographystyle{abbrvnat}
\bibliography{bibliography}



\end{document}